\documentclass{elsart}
\journal{Nuclear Physics A}
\usepackage{amstext,amssymb,graphicx}
\begin{document}
\begin{frontmatter}
\title{Low-lying dipole response in the Relativistic Quasiparticle
  Time Blocking Approximation and its influence on neutron capture
  cross sections}
\author[gsi,obninsk,fias]{E. Litvinova},
\author[gsi,tu]{H. P. Loens},
\author[gsi,fias,tu]{K. Langanke},
\author[gsi]{G. Mart\'inez-Pinedo},
\author[basel]{T. Rauscher},
\author[tum]{P. Ring},
\author[basel]{F.-K. Thielemann} and
\author[peter]{V. Tselyaev}
\address[gsi]{GSI Helmholtzzentrum f{\"u}r Schwerionenforschung,
  Planckstr. 1, 64291 Darmstadt, Germany} 
\address[obninsk]{Institute of Physics and Power Engineering, Pl. Bondarenko1,
249033 Obninsk, Russia} 
\address[fias]{Frankfurt Institute for Advanced Studies, Ruth-Moufang Str. 1,
60438 Frankfurt, Germany}
\address[tu]{Technische Univerist{\"a}t Darmstadt, Institut f{\"u}r
  Kernphysik, Schlossgartenstr. 9, 64289 Darmstadt, Germany}
\address[basel]{Department of Physics, University of Basel,
  Klingelbergstr. 82, 4056 Basel, Switzerland} 
\address[tum]{Physik-Department, Technische Universit{\"a}t
  M{\"u}nchen, 85748 Garching, Germany} 
\address[peter]{Nuclear Physics Department, V. A. Fock Institute of
  Physics, St. Petersburg State University, 198504 St. Petersburg,
  Russia} 
\begin{abstract}
  We have computed dipole strength distributions for nickel and tin
  isotopes within the Relativistic Quasiparticle Time Blocking
  approximation (RQTBA). These calculations provide a good description
  of data, including the neutron-rich tin isotopes $^{130,132}$Sn. The
  resulting dipole strengths have been implemented in Hauser-Feshbach
  calculations of astrophysical neutron capture rates relevant for
  r-process nucleosynthesis studies. The RQTBA calculations show the
  presence of enhanced dipole strength at energies around the neutron
  threshold for neutron rich nuclei. The computed neutron capture
  rates are sensitive to the fine structure of the low lying dipole
  strength, which emphasizes the importance of a reliable knowledge of
  this excitation mode.
\end{abstract}
\maketitle
\end{frontmatter}
\section{Introduction and motivation}

Neutron capture cross sections are one of the essential nuclear inputs
for studies of r-process nucleosynthesis. As this astrophysical
process, which produces about half of the heavy elements in the
universe, runs through nuclei with extreme neutron excesses, most of
the needed cross sections are not experimentally known and have to be
modelled.  Such estimates are usually done on the basis of the
statistical Hauser-Feshbach
\cite{Hauser:1952,Holmes.Woosley.ea:1976,Cowan.Thielemann.Truran:1991,Rauscher.Thielemann.Kratz:1997}
model and require, as one of the important ingredients, the modelling
of the electromagnetic transition strength from nuclear states above
the neutron threshold. Often dipole transitions contribute dominantly
to the cross sections. In the absence of experimental data, these
dipole strength distributions for neutron-rich nuclei have been
conventionally described by Lorentzians with an energy- or
temperature-dependent width
\cite{McCullagh:1981,RIPL2BOOK,Kopecky:1991} and parameters derived
from data for stable nuclei. Obviously deviations from the
parametrized Lorentzian at energies around the neutron threshold can
lead to significant modifications of the neutron capture cross
sections. As noted first by Goriely \cite{Goriely:1998}, such effects
can arise from the so-called pygmy resonance which leads to an
enhancement of the low-lying E1 strength, potentially in the
astrophysically important energy range. In fact, an enhancement of the
E1 strength at excitation energies $E_x$ below 10~MeV has been
experimentally observed for neutron-rich nuclei relative to the one
for stable nuclei. Prominent examples are the tin isotopes where the
neutron-rich nuclei $^{129-132}$Sn exhibit a noticeable portion of the
total E1 strength at $E_x < 10$~MeV, in contrast to the stable
isotopes $^{116,120}$Sn \cite{Klimkiewicz:2007}. Theoretically these
tendencies are well reproduced by calculations based on the
Quasiparticle Random Phase Approximation (QRPA) on top of
the Hartree-Fock plus Bardeen-Cooper-Schrieffer (HF-BCS) \cite{Goriely.Khan:2002}
or of the Hartree-Fock-Bogoliubov (HFB) \cite{Goriely:2004} approaches
and by the relativistic QRPA (RQRPA) calculations \cite{Paar:2003} and are
usually explained as so-called pygmy resonances in which the excess
neutrons oscillate against the isospin-saturated core. Significant
progress has been achieved very recently by extending the RQRPA by the
quasiparticle-phonon coupling model using the Relativistic
Quasiparticle Time Blocking Approximation (RQTBA). As it has been
demonstrated in \cite{Litvinova.Ring.Tselyaev:2008}, this extended
approach reproduces the experimental E1 strength distributions in
general very well, inluding the observed low-lying strength
It has been also found that isotopic dependence of the excitation energy
and the integrated strength of the pygmy dipole resonance agrees well 
with previous theoretical studies \cite{PVKC.07,TL.08}.
Ref.~\cite{Litvinova.Ring.ea:2008} has shown that once the
theoretical strength distributions are corrected for the experimental
detector response and in particular for the fact that Coulomb
dissociation experiments can only access strength above the neutron
threshold, the data for $^{130}$Sn and
$^{132}$Sn~\cite{Adrich.Klimkiewicz.ea:2005} are nicely
reproduced. This motivates us to use this model to calculate the E1
strength distributions for the chains of tin and nickel isotopes,
including nuclei close and on the r-process path for which data do not
exist, yet. The focus of our studies is on the effect of enhanced
low-lying dipole strength  on neutron capture cross sections, in
particular for neutron-rich nuclei.  Our calculations are based on the
statistical model which we briefly describe in the next section. In
section 3 we present the RQTBA dipole strength functions and compare
them with standard Lorentzian
parametrizations~\cite{Cowan.Thielemann.Truran:1991} and with the QRPA
calculations of Ref.~\cite{Goriely.Khan:2002}. The impact of
different dipole strength distributions on neutron-capture rates
for the nickel and tin isotopes is also discussed.

\section{Formalism}

The Hauser-Feshbach expression for the cross section of an
$(n,\gamma)$ reaction proceeding from the target nucleus $i$ in the
state $\mu$ with spin $J_i^\mu$ and parity $\pi_i^\mu$ to a final
state $\nu$ with spin $J_m^\nu$ and parity $\pi_m^\nu$ in the residual
nucleus $m$ via a compound state with excitation energy $E$, spin $J$,
and parity $\pi$ is given by (see also
\cite{Holmes.Woosley.ea:1976,Cowan.Thielemann.Truran:1991,Rauscher.Thielemann:2000}) 
\begin{equation}
 \sigma ^{\mu \nu}_{(n,\gamma)}(E_{i,n}) = \frac{\pi
   \hbar^2}{2M_{i,n}E_{i,n}} \frac{1}{(2J^{\mu}_i + 1)(2J_n+1)}
 \sum_{J,\pi} (2J+1)  
 \frac{T^{\mu}_n T^{\nu}
 _{\gamma}}{T_{\text{tot}}},
 \label{Hauser}
\end{equation}
where $E_{i,n}$ and $M_{i,n}$ are the center-of-mass energy and the
reduced mass for the initial system. $J_n=1/2$ is the neutron spin.
For $(n,\gamma)$ reactions the residual and the compound nucleus are
the same.  The transmission coefficients
$T_n^\mu=T_n(E,J,\pi;E_i^\mu,J_i^\mu,\pi_i^\mu)$,
$T_{\gamma}^\nu=T_\gamma (E,J,\pi;E_m^\nu,J_m^\nu,\pi_m^\nu)$ describe
the transitions from the compound state - characterized by $(E,J,\pi)$
- to the initial and final state, respectively. The sum of the
transmission coefficients over all possible exit-channels is given by
$T_\mathrm{tot}$. It is possible to derive the laboratory cross
section (the target is in its ground state, e.g. $\mu = 0$) from
Eq. (\ref{Hauser}) by summing over all possible final states $\nu$ in
the residual nucleus. Therefore it is convenient to introduce
transmission coefficients which only depend on the compound quantum
numbers:

\begin{equation}
    T_{\gamma}(E,J,\pi) = \sum_{\nu}
    T_{\gamma}(E,J,\pi;E_m^\nu,J_m^\nu,\pi_m^\nu). 
    \label{avgT}
\end{equation}
Generally the discrete energy spectrum is only partially known and
consequently above a certain excitation energy a level density
description is applied. In that case the sum in
Eq. (\ref{avgT}) translates into a discrete sum plus an integration
over the level density
\cite{Holmes.Woosley.ea:1976,Cowan.Thielemann.Truran:1991,Rauscher.Thielemann:2000}:

\begin{eqnarray}
    T_{\gamma}(E,J,\pi) & = &\sum_{\nu=0}^{\kappa}
    T_{\gamma}(E,J,\pi;E_m^\nu,J_m^\nu,\pi_m^\nu) \\ \nonumber 
    & + & \int _{E^{\kappa}_m} ^{E}\sum_{J_m,\pi_m}
    T_{\gamma}(E,J,\pi;E_m,J_m,\pi_m) \rho(E_m,J_m,\pi_m) dE_m. 
    \label{intT}
\end{eqnarray}
where $E_m^\kappa$ is the excitation energy of the last experimentally 
known state,
$\kappa$, in the nucleus $m$. 

In the calculation of the transmission coefficient we closely follow
the formalism as described in
\cite{Cowan.Thielemann.Truran:1991,Thielemann:1987neu}. In particular
we consider parity-dependent level densities for initial, compound and
final states, here following \cite{Loens.Langanke.ea:2008}. The
parity-dependent level densities were taken from Hilaire and Goriely
as calculated within a combinatorial approach on top of HFB single
particle energies~\cite{Hilaire.Goriely:2006}.  We note that using a
standard backshifted Fermi gas level
density~\cite{Rauscher.Thielemann.Kratz:1997,Cowan.Thielemann.Truran:1991}
with a parity-dependence as introduced in
\cite{Mocelj.Rauscher.ea:2007} gives quite similar results.


The $\gamma$-transmission coefficient for the decay to a state $\nu$
involves a sum over all possible photon multipolarities.  For a given
multipolarity, XL, and photon energy $E_{\gamma}=E-E^\nu$, the
transmission coefficient is related with the $\gamma$-strength
function, $f_{\text{XL}}$, via

\begin{equation} 
T_{\text{XL}}(E,J,\pi,E^\nu,J^\nu,\pi^\nu) =
    T_{\text{XL}}(E,E_{\gamma}) = 2\pi E^{2L+1}_{\gamma}
     f_{\text{XL}}(E,E_{\gamma}).  \label{t_e1} 
\end{equation} 

In this paper we mainly focus on E1 transitions, although the M1
contributions have been calculated as well using the single-particle
approach \cite{Holmes.Woosley.ea:1976,Blatt.Weisskopf:1952}. Both
multipolarities usually dominate for neutron capture reactions at
astrophysical energies.

In general, the strength function depends on the initial and final states.
However, in practical applications, one usually adopts the Brink-Axel
hypothesis which assumes that the strength function is independent of the
detailed structure of the initial state and consequently depends only on
$E_\gamma$ and not on $E$~\cite{Bartholomew:1973,Brink:1955,Axel:1968}.
Assuming that the photoabsorption process does not depend on the spin of the
final state, it is possible to relate the cross section,
$\sigma_{\text{XL}}$, to the $\gamma$-strength function:

\begin{equation}
f_{\text{XL}} (E, E_\gamma) = 
f_{\text{XL}} (E_\gamma) = 
\frac{1}{2L+1}
    \frac{\sigma_{\text{XL}}(E_\gamma)}{(\pi\hbar c)^2} \cdot E_{\gamma}^{-2L+1}. 
    \label{f_up}
\end{equation}




According to Eq. (\ref{f_up}), the $\gamma$-strength function and the
absorption cross section for E1 transitions are connected via 

\begin{equation}
    \sigma_{\text{E1}}(E_\gamma) = 3\pi^2 (\hbar c)^2 E_{\gamma}
    f_{\text{E1}} (E_\gamma).
    \label{friedel_f}
\end{equation}



In statistical model calculations one usually assumes that the E1
strength distribution can be approximated by a Lorentzian adjusted to
reproduce the position and width of the giant
dipole resonance
(GDR)~\cite{Holmes.Woosley.ea:1976,Thielemann:1987neu,Cowan.Thielemann.Truran:1991,RIPL2BOOK,Kopecky:1991}: 

\begin{equation}
    f_{\text{E1}}(E_{\gamma}) = \frac{4}{3\pi} \frac{e^2}{\hbar c}
    \frac{1}{mc^2} \frac{NZ}{A} \frac{\Gamma
      E_{\gamma}}{(E_{\text{E1}}^2 - E_{\gamma}^2)^2 +
      (E_{\gamma}\Gamma)^2} (1+\chi)
    \label{e1_lor}
\end{equation}
with $N$, $Z$ and $A=N+Z$ being the neutron, proton and nucleon
numbers of the compound nucleus. Here $\chi\approx0.2$ is a factor
that accounts for the neutron-proton exchange contribution (see e.g.,
\cite{Lipparini:1989}). The parameters $E_{\text{E1}}$ and $\Gamma$
are obtained by adjustment to data for the giant dipole resonance (see
e.g., \cite{Dietrich:1988}) or, if data are not available, from
theoretical model predictions or
systematics~\cite{Cowan.Thielemann.Truran:1991}. 
The Lorentzian approach is known to be inaccurate in the
low-energy region as the low energy tail of the Lorentzian usually
overestimates the strength at these energies. In practical applications
one therefore modifies the width of the GDR, $\Gamma$, to account
for these deficiencies. Various treatments have been developed over
the past decades (for an overview see \cite{RIPL2BOOK} and references therein).
Here we follow \cite{McCullagh:1981} and define
an energy-dependent width parameter as:
$\Gamma (E_\gamma) = \Gamma \sqrt{E_\gamma/E_{\text{E1}}}$.

We have calculated RQTBA E1 strength functions as outlined in
\cite{Litvinova.Ring.Tselyaev:2008} and used them in combination with Eqs.
(\ref{t_e1},\ref{friedel_f}) to obtain transmission coefficients
for the E1 transitions. 
In the RQTBA, excitations in even-even nuclei are determined by
the nuclear response function $R(\omega)$ whose matrix elements
obey the Bethe-Salpeter equation (BSE) of the following form
\cite{Litvinova.Ring.Tselyaev:2008}:
\begin{eqnarray}
R_{k_{1}k_{4},k_{2}k_{3}}^{\eta\eta^{\prime}}(\omega) &=& \tilde{R}_{k_{1}k_{2}%
}^{(0)\eta}(\omega)\delta_{k_{1}k_{3}}\delta_{k_{2}k_{4}}\delta_{\eta
\eta^{\prime}} \nonumber \\
&+& \tilde{R}_{k_{1}k_{2}}^{(0)\eta}(\omega)\sum\limits_{k_{5}%
k_{6}}\sum\limits_{\eta^{\prime\prime}}{\bar{W}}_{k_{1}k_{6}%
,k_{2}k_{5}}^{\eta\eta^{\prime\prime}}(\omega)R_{k_{5}k_{4},k_{6}k_{3}}%
^{\eta^{\prime\prime}\eta^{\prime}}(\omega),
\label{respdir}%
\end{eqnarray}
where indices $k_i$ run over single-particle quantum numbers
including states in the Dirac sea and indices $\eta,
\eta^{\prime}, \eta^{\prime\prime}$ numerate forward (+) and
backward ($-$) components in the quasiparticle space. Pairing
correlations are treated in the BCS approximation. The quantity
$\tilde{R}^{(0)}(\omega)$ describes free propagation of two
quasiparticles in the mean field and matrix elements of the
amplitude $\bar{W}$ read:
\begin{equation}
{\bar{W}}_{k_{1}k_{4},k_{2}k_{3}}^{\eta\eta^{\prime}}(\omega)=\tilde{V}%
_{k_{1}k_{4},k_{2}k_{3}}^{\eta\eta^{\prime}}+\Bigl(\Phi_{k_{1}k_{4},k_{2}%
k_{3}}^{\eta}(\omega)-\Phi_{k_{1}k_{4},k_{2}k_{3}}^{\eta}(0)\Bigr)\delta
_{\eta\eta^{\prime}}.
\label{W-omega}%
\end{equation}
In Eq. (\ref{W-omega}) $\tilde V$ is the static part of the amplitude
${\bar{W}}$ based on a one-meson exchange interaction with a
non-linear self-coupling between the mesons and determined by the
relativistic energy functional with the parameter set NL3
\cite{Lalazissis.Koenig.Ring:1997}. $\Phi (\omega)$ is the dynamical
part of the interaction amplitude responsible for the particle-phonon
coupling. It has been calculated in the quasiparticle time blocking
approximation, see \cite{Litvinova.Ring.Tselyaev:2008} for details.

To describe the observed spectrum of an excited nucleus in a weak
external electromagnetic field $P$, one needs to calculate the
strength function which is a convolution of the response function of
Eq. (\ref{respdir}) with matrix elements of the external field
operator $P_{k_{1}k_{2}}^{\eta}$:
\begin{equation}
S(E_\gamma) = -\frac{1}{2\pi}\lim\limits_{\Delta\rightarrow+0}Im\
\sum\limits_{k_{1}k_{2}k_{3}k_{4}}\sum\limits_{\eta\eta^{\prime}}
P_{k_{1}k_{2}}^{\eta\ast}R_{k_{1}k_{4},k_{2}k_{3}}^{\eta\eta^{\prime}}
(E_\gamma + i\Delta)P_{k_{3}k_{4}}^{\eta^{\prime}}.
\label{strf}%
\end{equation}
In the calculations a small finite imaginary part $\Delta$ of the
energy variable is introduced for convenience in order to obtain a
more smoothed envelope of the spectrum. This parameter has the
meaning of an additional artificial width for each excitation.
This width emulates effectively contributions from configurations
which are not taken into account explicitly in our approach.

The dipole photo-absorption cross section,
\begin{equation}
\sigma_{E1}(E_\gamma)={\frac{{16\pi^{3}e^{2}}}{{9\hbar c}}}E_\gamma
S_{E1}(E_\gamma), 
\end{equation}
is expressed through the microscopically computed strength
function of Eq.~(\ref{strf}) and determines the dipole
$\gamma$-strength function of Eq. (\ref{friedel_f}) which is used
for the statistical model calculations.

\section{Results}

We have calculated statistical model cross sections for $(n,\gamma)$
reactions on tin and nickel isotopes using four different models to
describe the dipole strength function. Our default model (model A) is
the RQTBA which, as demonstrated
in~\cite{Litvinova.Ring.Tselyaev:2008}, reproduces experimental photo
absorption cross sections very well, including the low-energy cross
sections, which is of special interest for the discussion here. To
account effectively for the residual effect of higher configurations
beyond RQTBA, a 200 keV imaginary part has been included in the energy
variable of Eq.~(\ref{strf}). As the model is currently restricted to
even-even nuclei, we will in the following only discuss neutron
capture cross sections for odd-$A$ nickel and tin isotopes.  Following
Refs. \cite{Holmes.Woosley.ea:1976,Cowan.Thielemann.Truran:1991,%
  Thielemann:1987neu} we describe the dipole strength by a Lorentzian
form of Eq. (\ref{e1_lor}) with the energy dependent width parameter
as defined above.  The centroid $E_{\text{E1}}$ and total strength are
adjusted to the RQTBA strength function (model B). Although the width
parameter is taken from Ref. \cite{Thielemann:1983}, it gives a good
reproduction of the RQTBA strength function in the GDR region.  A
comparison between models A and B allows to disentangle the influence
of the low-energy strength on the neutron capture cross
sections. Finally, we have performed calculations using the Lorentzian
form of Eq.~(\ref{e1_lor}) but with the strength parameters determined
as described in Ref.~\cite{Cowan.Thielemann.Truran:1991} (Model C) and
calculations adopting the dipole strength functions of
Ref. \cite{Goriely.Khan:2002}, which have been derived by a QRPA
approach on top of a Hartree-Fock-BCS model (model D).

\begin{figure}[htb]
  \includegraphics[width=0.9\columnwidth]{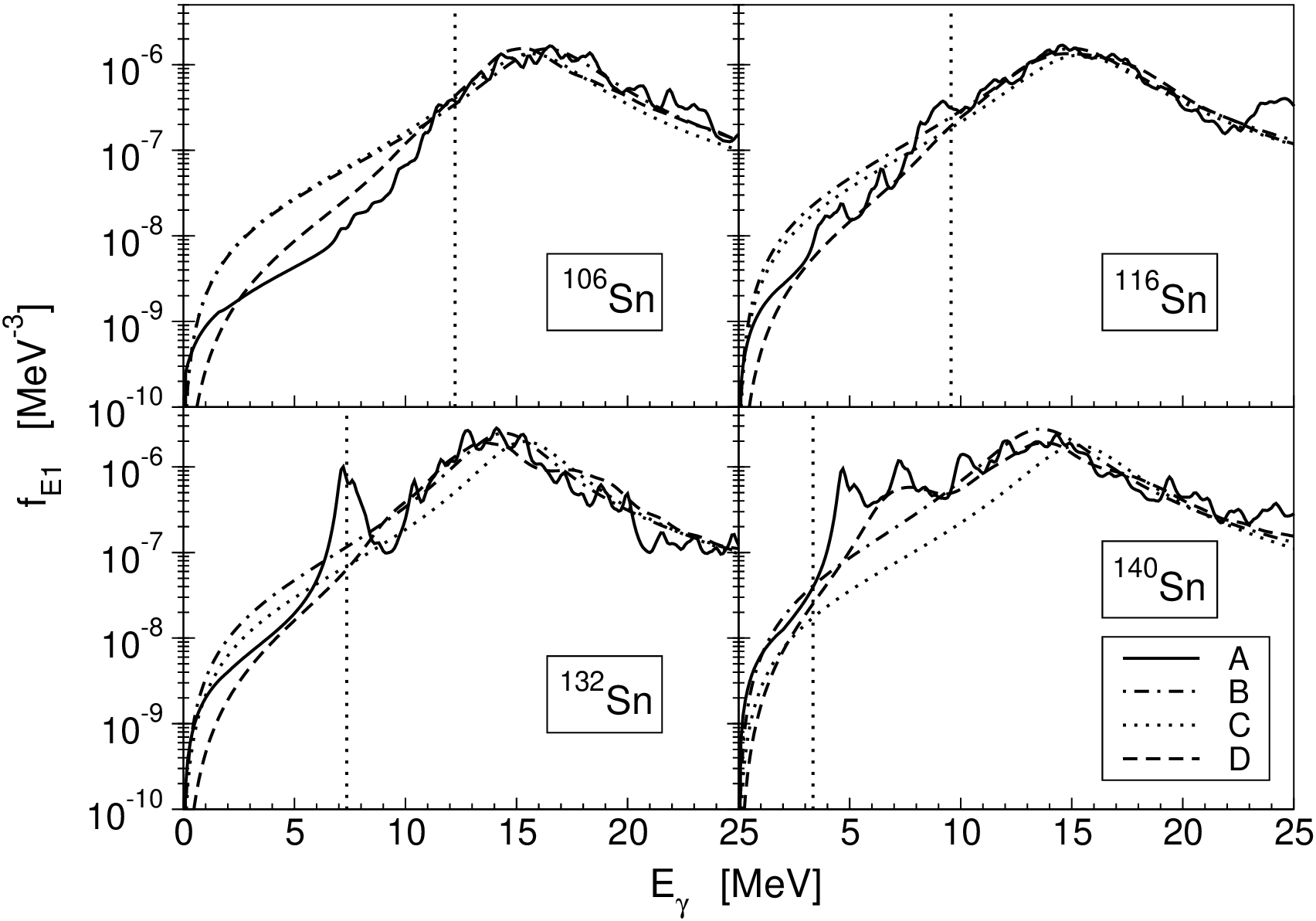}
  \caption{$\gamma$ strength functions for selected tin isotopes as
    calculated by the RQTBA approach (solid; model A)
    \cite{Litvinova.Ring.Tselyaev:2008} and the QRPA model of Goriely
    and Khan \cite{Goriely.Khan:2002} (dashed; model D). The dotted
    curves show the Lorentzian parametrization of the strength as
    proposed in \cite{Cowan.Thielemann.Truran:1991} (model C) and the
    dash-dotted curves correspond to a Lorentzian with the width
    taken from \cite{Cowan.Thielemann.Truran:1991}, the centroid taken
    from the RQTBA results and the total strength adjusted to the RQTBA
    results, as explained in the text (model B). The dotted vertical
    lines indicate the (experimental) neutron threshold
    energies.\label{fE1}}
\end{figure}

Fig. \ref{fE1} shows the dipole $\gamma$ strength functions for the
four different models and selected tin isotopes.  We note that models
A and D reproduce the known
photoabsorption cross section data for tin isotopes quite well
\cite{Litvinova.Ring.Tselyaev:2008,Litvinova.Ring.ea:2008}.  While the parametrized strength
function of Ref. \cite{Cowan.Thielemann.Truran:1991} agrees nicely
with the microscopic $^{116}$Sn strength functions, it predicts a
centroid energy which is somewhat higher for the neutron-rich tin
isotope $^{132}$Sn than obtained in the other models.

Comparing the various model predictions it is most striking that the
microscopic model predicts quite noticeable nuclear structure effects
which result in a rather strong fragmentation of the strength
function. This is mainly due to the quasiparticle-phonon coupling
explicitly considered in the RQTBA
approach~\cite{Litvinova.Ring.Tselyaev:2008}. Comparing the
microscopic strength functions (model A) with the Lorentzian
approximation of model B, we note that for most nuclei the
parametrization overestimates the microscopic strength function at low
energies, say $E < 5$~MeV. This indicates that the empirical
energy-dependence of the width parameter \cite{McCullagh:1981} is not
supported by the microscopic model. This conclusion is also supported
by the QRPA results which are generally slightly smaller than the
RQTBA results at the lowest energies. The standard parametrization
utilized in \cite{Cowan.Thielemann.Truran:1991} agrees quite well with
the model B parametrization. However, with increasing neutron excess
this model predicts the centroid at higher energies than the other
models. Consequently, the $\gamma$ strength function for $^{140}$Sn in
model C is smaller at low energies than in the other models.

As the RQTBA model includes phonon couplings which are absent in the
QRPA model it resolves more nuclear structure details. This can lead
to fluctuations in the $\gamma$ strength function around the neutron
threshold and below.  An example is given by the nucleus $^{116}$Sn
(see Fig. \ref{fE1}) where the RQTBA model predicts a mild enhancement
of the strength just at the neutron threshold which is absent in the
QRPA model and, of course, in the parametrizations of models B and C.
As the neutron separation energies move to lower energies with
increasing neutron excess within a chain of isotopes, low-lying E1
strength as observed experimentally for the tin isotopes
$^{130,132}$Sn and found within our microscopic RQTBA approach
for neutron-rich nuclei can influence the $\gamma$ strength
function around the neutron threshold significantly. A prominent
example is $^{132}$Sn where the RQTBA approach predicts a strong
enhancement of the E1 strength around the neutron separation energy of
$E=7.3$~MeV. Such an increased strength is not predicted by the other
models used in this paper (the relativistic QRPA calculations of
Ref.~\cite{Paar:2003} predict also additional strength at low
energies) and, as we will see below, has a noticeable effect on the
neutron capture cross sections. For $^{140}$Sn the RQTBA model also
predicts a strong fragmentation of the $\gamma$ strength at low
energies $E \gtrsim 5$~MeV with enhanced strength compared to the strengths
of the other models. However, for this nucleus the neutron threshold
is already quite low ($E_{\text{th}} = 3.5$~MeV) and this enhancement
will only have modest effect on the cross sections.

\begin{figure}[htb]
  \includegraphics[width=0.9\columnwidth]{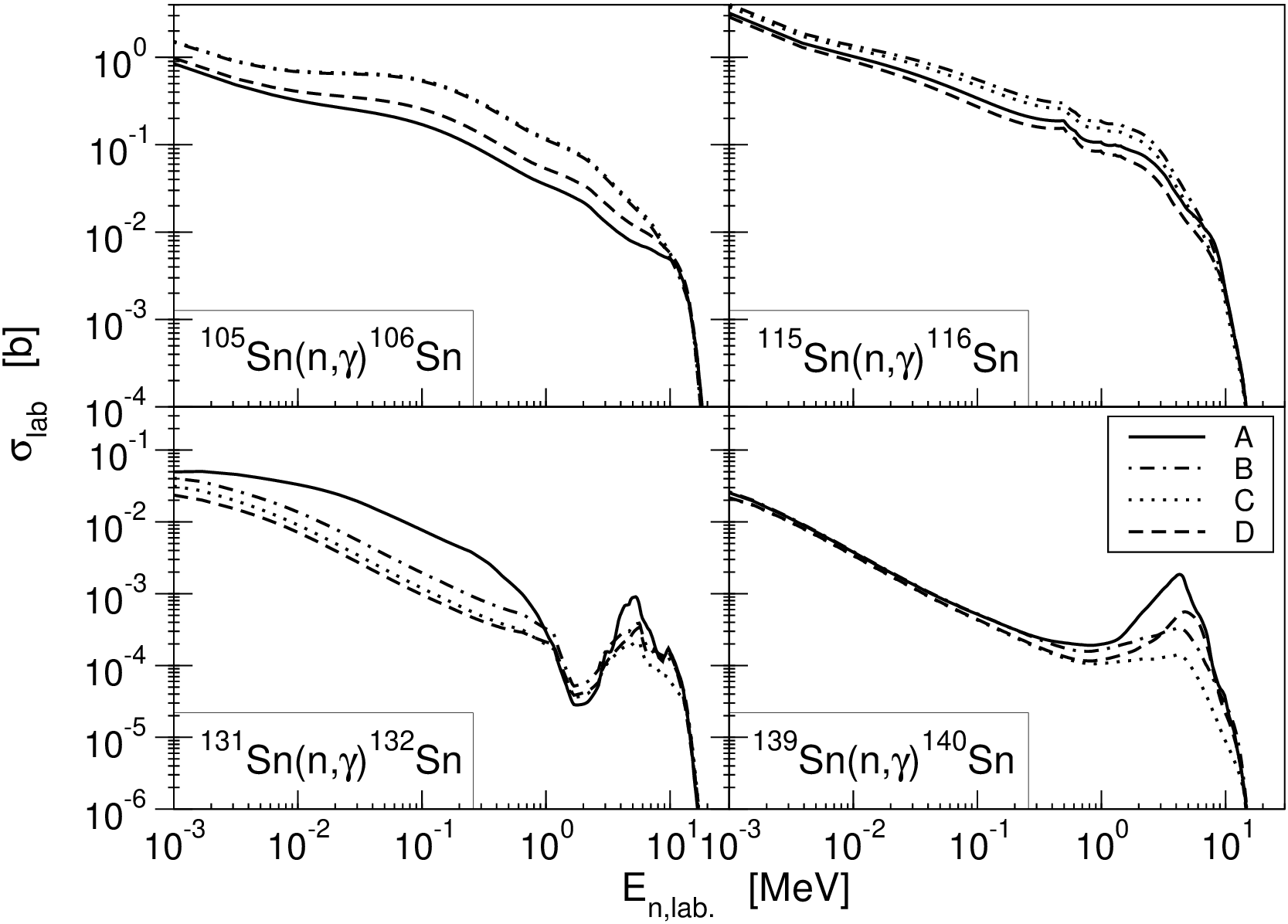}
  \caption{Cross sections for $(n,\gamma)$-reactions on various tin
    isotopes based on E1 strength functions as derived in different
    models, see Fig. \ref{fE1} and text.\label{sn-cs}}
\end{figure}

By inserting the $\gamma$ strength functions into Eqs. (1) and (4) we
have calculated cross sections for neutron capture on the ground
states of the isotopes $^{105,115,131,139}$Sn arising solely from
dipole transitions. The results are shown in Fig.~\ref{sn-cs} for
neutron energies of up to 20 MeV (note that the maximum neutron
energies reached in the r-process discussed below are of the order of
$80-100$ keV). In order to understand the impact of various magnitudes
of $\gamma$ strengths it is necessary to realize what the relevant
$\gamma$-ray energies are. It is well known that capture reactions
proceeding via the compound reaction mechanism result in a
$\gamma$-ray cascade with a dominant contribution coming from $\gamma$
rays with only a few MeV, below the neutron separation energy
$S_n$. Recently, Ref.~\cite{Rauscher:2008}, assuming the validity of
the Brink-Axel hypothesis and a Lorentzian parametrization of the
dipole strength, has shown that this also holds for neutron-rich targets 
with dominant E1 strength leading to levels at 3-4 MeV below
$S_n$
unless the density of levels accessible by E1 transition is
low. In the latter case, $\gamma$ transitions with the maximally
possible energy of $S_n+E_n$ may dominate. This nicely explains the
dependences we find in our calculations. For the cases of neutron
captures on $^{105}$Sn and $^{115}$Sn the dominating $\gamma$
transitions display an energy of $E_\gamma \approx 3+E_n$ MeV 
with dominant E1 transition endpoints at $\approx 3$ MeV
below $S_{n}$.
Scanning through $E_n$, we can nicely see in Figs. \ref{fE1},\ref{sn-cs} how the
larger strength functions of the two parametrized descriptions (models
B and C) result in larger cross sections than those obtained with the
microscopic approaches RQTBA and QRPA (models A and D), until the
predictions converge close to $S_n$. The RQTBA model predicts a strong
fragmentation of the dipole strength at low energies. However, this is
not seen in the cross sections because the relevant range of available
$\gamma$ energies averages and smoothes the impact of the
fluctuations.  The small kink appearing in the $^{115}$Sn cross
section at $E_n=0.5$ MeV is due to the opening of the $(n,p)$ channel.

Consistent with~\cite{Rauscher:2008}, we find a different behavior in
the cross sections of $^{131}$Sn. The low level density of the doubly
magic compound (and final) nucleus $^{132}$Sn causes the dominance of
$\gamma$ transitions directly to the ground state, with
$E_\gamma=S_n+E_n$.  Up to $E_n<1$~MeV the
influence of the RQTBA pygmy peak explains why the RQTBA yields the by
far largest cross sections up to that energy. Above that energy, the
RQTBA $\gamma$ strength drops below the strengths predicted by the
other models and consequently the resulting cross section becomes the
smallest. After another $2-3$ MeV (note that Fig. 1 uses a linear and
Fig. 2 a logarithmic energy scale), the RQTBA predicts a larger
$\gamma$ strength again and consequently its respective cross section
becomes the largest again. Interestingly, a second effect is overlaid
on top of the general strength function behavior and giving rise to
the decrease in cross sections found for all models around 2 MeV
neutron energy. The level density
description~\cite{Hilaire.Goriely:2006} predicts a 
relatively low density of negative parity states at excitation
energies around 9 MeV in $^{132}$Sn. In consequence, magnetic dipole
transitions to the low-lying positive parity states 
dominate (not shown in Fig. 2). This is
an example for the importance of the complete inclusion of
parity-dependent level densities.

Except for the parity effect, the behavior of the
$^{139}$Sn$(n,\gamma$) cross section can be explained similarly to the
one of $^{131}$Sn. The level density below $S_n$ is very low for such
a neutron-rich nucleus and therefore the relevant $\gamma$ energies
are $E_\gamma=S_n+E_n$. The strong pygmy peak predicted by the RQTBA
is also clearly seen in the cross section. Smaller fluctuations are
dampened and the general strong decrease in the cross section at the
high energy end is typical for Hauser-Feshbach cross sections, as in
the other reactions considered here. In addition, for $^{139}$Sn we
find
that the M1 contribution
to the capture cross section exceeds the dipole part at low neutron
energies (say $E_n < 100$ keV).  Thus, at these energies the total
cross sections are less affected by the differences in the dipole
strength functions. However, our statistical model code uses a rather
simple description of M1 transitions based on a single particle
treatment~\cite{Holmes.Woosley.ea:1976,Blatt.Weisskopf:1952}. A more
appropriate treatment, which includes the concentration of the M1
strength around $E=7$--10~MeV due to spin excitations and in addition
considers the pronounced orbital contribution seen in deformed nuclei
(scissors mode~\cite{Richter:1995}) is called for. 

\begin{figure}[htb]
  \includegraphics[width=0.9\columnwidth]{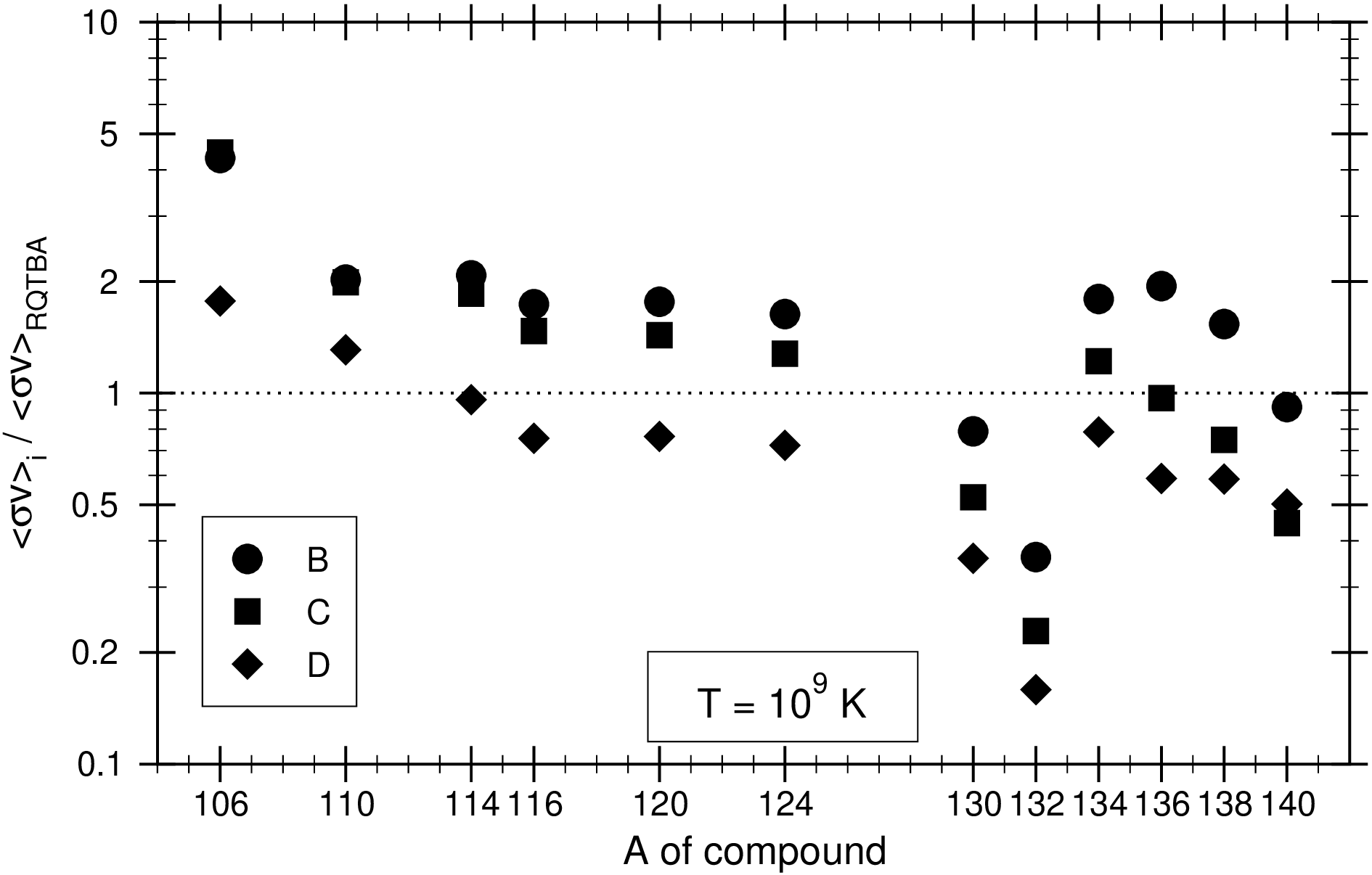}
  \caption{Ratios of the stellar neutron capture rates at a
    temperature $T=10^9$~K for tin isotopes as calculated within
    models B, C, and D relative to those obtained in model A
    (RQTBA). All the calculations consider only E1 gamma
    transitions. For descriptions of the models, see
    text.\label{sn-ratios}}
\end{figure}

The classical r-process operates at a typical stellar temperature of
order $10^9$~K~\cite{Cowan.Thielemann.Truran:1991}. To estimate which
impact the enhanced low-energy dipole strength could have on neutron
capture rates under astrophysical conditions, we have calculated the
stellar neutron capture rate $\langle \sigma v \rangle$, where
$\sigma$ is the neutron capture cross section, $v$ is the relative
velocity of the fusing nuclei in the initial channel and the symbol
$\langle ... \rangle$ indicates proper averaging over the
Maxwell-Boltzmann velocity distributions of the two nuclei at the
stellar temperature and accounting for the thermal population of
excited states. In Fig. \ref{sn-ratios} we compare ratios of neutron
capture rates as obtained in the four different models at a
temperature of $T = 10^9$ K.  The capture rates have been computed
neglecting M1 transitions due to their rather simple treatment in our
statistical code (see paragraph above). We observe that rates
calculated with the parametrized E1 strengths are usually larger than
our RQTBA results with the noticeable exception for the isotopes
$^{130,132}$Sn (capturing the neutron on $^{129,131}$Sn).  With the
additional observation that the other microscopically calculated
strength functions (QRPA, model D) generally predicts smaller rates
than the parametrization models one might conclude that the
parametrization of the dipole strength functions with the energy
dependence as defined above seems to predict too large strength at low
gamma energies (except for the most neutron-rich tin isotope,
$^{140}$Sn, considered in this work) and hence overestimates the
rates, compared to microscopic models. Interestingly, we do not
observe by comparison of the four models that the low-lying strength
is more important in neutron-rich nuclei as, for example, the
parametrized strength of model B exceeds the RQTBA results by about
the same amount in the neutron deficient and stable tin isotopes as it
does in the very neutron-rich ones ($^{133-137}$Sn).


\begin{figure}[htb]
  \includegraphics[width=0.9\columnwidth]{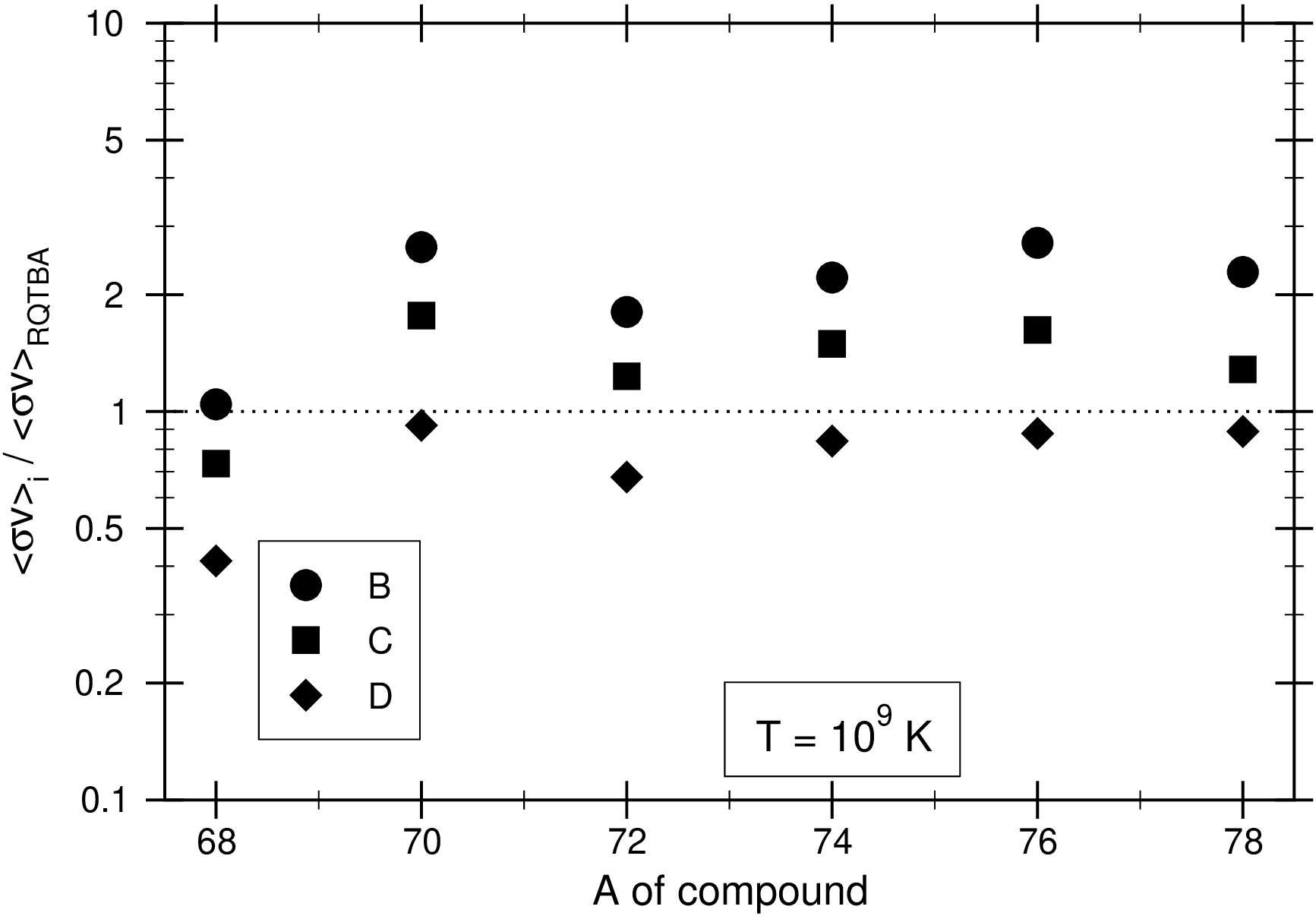}
  \caption{Ratios of the stellar neutron capture rates at a
    temperature $T=10^9$~K for nickel isotopes as calculated within
    models B, C, and D relative to those obtained in model A
    (RQTBA). All the calculations consider only E1 gamma
    transitions. For descriptions of the models, see
    text.\label{ni-ratios}}
\end{figure}

We have also performed calculations of dipole strength functions,
cross sections and neutron capture rates at $T=10^9$ K for the even
neutron-rich nickel isotopes $^{68-78}$Ni. A comparison of the rates
obtained in our four different models is shown in
Fig. \ref{ni-ratios}. The findings are quite similar to those for the
tin isotopes. For all isotopes the rates calculated with the
parametrized dipole strength (models B and C) exceeds those obtained
on the basis of the RQTBA strength functions, again pointing to a
non-adequate description of the low-lying strength by the energy
dependent width parameter as chosen
in~\cite{Cowan.Thielemann.Truran:1991}. Except for the capture on
$^{67}$Ni, the agreement between the two microscopic approaches 
(models A and D) is quite satisfactory. The Lorentzian parametrization
slightly overestimates the microscopic rates. We note, however, that
for the heaviest nickel nuclei the parametrization used within
Ref. \cite{Cowan.Thielemann.Truran:1991} predicts the centroid of the
giant dipole resonance at slightly higher energies. If corrected, the
downwards shift in energy would result in somewhat larger rates.

In general, our calculations for the capture rates on tin and nickel isotopes
show a rather close agreement between the models suggesting that
the enhanced low-lying dipole strength predicted by theoretical models
and experimentally observed for selected nuclei has only modest influence
on the neutron capture rates. The largest effect we find for the
capture on the tin isotopes $^{129,131}$Sn (caused by
low-lying dipole strength in $^{130,132}$Sn) where our microscopic RQTBA
model predicts noticeable enhanced strength just around the energies of
the neutron threshold, leading to an increase in the capture rate
by factors 2--5 compared to the other models considered here.

The coincidence that the enhanced E1 strength lies around the neutron
threshold in $^{132}$Sn and a large portion of the transitions goes to
the daughter ground state (due to suppression of excited states at
modest excitation energies in this double-magic nucleus) results in a
large enhancement of the reaction rate in the RQTBA model compared to
the other models.  As the RQTBA strength function for $^{132}$Sn
nicely agrees with data (see Fig. 1 of
Ref.~\cite{Litvinova.Ring.ea:2008}), this change in the capture rate
should be adopted in future r-process simulations.

\section{Conclusions}

We have obtained E1 strength functions for nickel and tin isotopes
based on the microscopic RQTBA approach which has been proven
previously to give a good description of data, including those for the
neutron-rich tin isotopes $^{130,132}$Sn. These data show E1 strengths
at energies around the neutron threshold which are enhanced compared
to the expectations from a Lorentzian parametrization of the giant
resonance.  As it has been pointed out before that such enhanced
dipole strengths could have noticeable effects on neutron capture
rates for r-process nuclei, we have calculated the respective capture
cross sections for the relevant nickel and tin isotopes ($^{67-77}$Ni
and $^{129-139}$Sn) and have compared the RQTBA results with rates
obtained either within the non-relativistic QRPA on the basis of
Hartree-Fock plus BCS calculations or by empirical parametrizations.
Usually we find rather good agreement between the two microscopic
models. Noticeable exceptions are $^{129,131}$Sn, and in particular
the capture rates to $^{132}$Sn are noticeably larger within the RQTBA
model than in all other models, which is caused by the enhanced dipole
strength in this nucleus at energies above the neutron threshold.  We
note, however, that our calculation of the capture rates are based on
the Brink-Axel hypothesis, which is common in statistical model
evaluations of neutron capture rates for r-process applications.
Within this hypothesis an enhanced dipole strength around the neutron
threshold translates to enhanced neutron capture cross sections only
if transitions to the ground state dominate. If the capture leads to
excited states instead the strength function at appropriate lower
energies is relevant. However, the transition strength for excited states
could be different than for the ground state at the moderate energies
relevant to neutron capture, invalidating the Brink-Axel
hypothesis. Consequently, tests of this hypothesis for
dipole transitions, similar to those performed for Gamow-Teller
transitions in Ref.~\cite{Langanke.Martinez-Pinedo:2000} are welcome.

Two final remarks are in order. At first, the RQTBA model predicts
fragmentation and low-lying strengths also in the neutron-rich
isotopes, however, not at the low energies close to the neutron
threshold. Whether further correlations beyond the RQTBA model will
further push strength to lower energies and hence might enhance the
dipole cross sections at low neutron energies is an open question and
must wait until appropriate nuclear models can be applied to such
nuclei. Secondly, we have also calculated the contributions of other
multipoles to the neutron capture rate and find that for the tin
isotopes beyond $^{132}$Sn, which have quite low neutron thresholds,
M1 transitions dominate in our calculations making the observed
differences in the dipole cross sections unimportant for the total
cross sections due to our use of parity-dependent level
densities~\cite{Loens.Langanke.ea:2008}. However, we note that our
statistical model code describes M1 transitions in the so-called
single-particle treatment. The calculation of the M1 transitions
should generally be improved by using M1 strength functions which
resemble the experimentally observed energy distribution with strong
spin contributions around excitation energies of order 7--10~MeV and
orbital contributions at low energies, including the pronounced
scissors mode~\cite{Richter:1995} in deformed nuclei at excitation
energies of $E \approx 2$--3~MeV, i.e.  around neutron threshold
energies in very neutron-rich nuclei.

\ack

This work is partly supported by the Deutsche Forschungsgemeinschaft
through contract SFB 634 ``Nuclear structure, nuclear astrophysics and
fundamental experiments at small momentum transfers at the
S-DALINAC''. T. Rauscher and F.-K. Thielemann are supported by the
Swiss National Foundation (grant 200020-122287). P. Ring is partly
supported by the DFG cluster of excellence ``Origin
and Structure of the Universe'' (www.universe-cluster.de).
E. Litvinova and V. Tselyaev gratefully acknowledge support from
the Russian Federal Education Agency Program, project  2.1.1/4779.


\end{document}